\documentclass[preprint]{aastex6}

\usepackage{graphicx}
\usepackage{subcaption}
\usepackage{amsmath}
\usepackage{cases}

\begin{document}

\title{Direct T$_e$ metallicity calibration of R23 in strong line emitters}

\author{Tianxing Jiang\altaffilmark{1,2,3}, Sangeeta Malhotra\altaffilmark{4}, 
James E. Rhoads\altaffilmark{4}, Huan Yang\altaffilmark{5}}
\altaffiltext{1}{School of Earth and Space Exploration, Arizona State University, Tempe, AZ 85287, USA; tianxing.jiang@asu.edu}
\altaffiltext{2}{LSSTC Data Science Fellow}
\altaffiltext{3}{Department of Astronomy, University of Maryland, College Park, MD 20742, USA}
\altaffiltext{4}{NASA Goddard Space Flight Center, Greenbelt, MD 20771, USA}
\altaffiltext{5}{Las Campanas Observatory, Carnegie Observatories, La Serena, Chile}

\begin{abstract}

The gas metallicity of galaxies is often estimated using strong emission lines such as the optical lines of [OIII] and [OII]. The most common measure is ``R23'', defined as ([OII]$\lambda$$\lambda$3726, 3729 + [OIII]$\lambda$$\lambda$4959,5007)/H$\beta$. Most calibrations for these strong-line metallicity indicators are for continuum selected galaxies. We report a new empirical calibration of R23 for extreme emission-line galaxies using a large sample of about 800 star-forming green pea galaxies with reliable T$_e$-based gas-phase metallicity measurements. This sample is assembled from Sloan Digital Sky Survey (SDSS) Data Release 13 with the equivalent width of the line [OIII]$\lambda$5007 $>$ 300 \AA\ or the equivalent width of the line H$\beta$ $>$ 100 \AA\ in the redshift range 0.011 $<$ z $<$ 0.411. For galaxies with strong emission lines and large ionization parameter (which manifests as log [OIII]$\lambda$$\lambda$4959,5007/[OII]$\lambda$$\lambda$3726,3729 $\geq$ 0.6), R23 monotonically increases with log(O/H) and the double-value degeneracy is broken. Our calibration provides metallicity estimates that are accurate to within $\sim$ 0.14 dex in this regime. Many previous R23 calibrations are found to have bias and large scatter for extreme emission-line galaxies. We give formulae and plots to directly convert R23 and [OIII]$\lambda$$\lambda$4959,5007/[OII]$\lambda$$\lambda$3726,3729 to log(O/H). Since green peas are best nearby analogs of high-redshift Lyman-$\alpha$ emitting galaxies, the new calibration offers a good way to estimate the metallicities of both extreme emission-line galaxies and high-redshift Lyman-$\alpha$ emitting galaxies. We also report on 15 galaxies with metallicities less than 1/12 solar, with the lowest metallicities being 12+log(O/H) = 7.25 and 7.26.

\end{abstract}


\section{Introduction}

In the galactic ecosystem, stars form from the collapse of gas clouds and fuse hydrogen and helium into heavy elements (metals); stars eject gas and metals into the interstellar medium by stellar feedback; cool gas in the circumgalactic and intergalactic medium flows into the galaxy; and gas enriched with metals in the galaxy can be transported into the intergalactic medium by galactic outflows. The fraction of gas that has been converted to heavy elements, which is often quantitatively characterized by ``metallicity", is key for understanding the star formation history and galactic chemical evolution. In addition, metallicity impacts the luminosity and color of the stellar light, the cooling of gas, and the amount of dust, which in turn determines the interstellar extinction. Robust metallicity measurement is the foundation for investigating mass-metallicity and mass-metallicity-SFR relations and their redshift evolution. 

The gas-phase oxygen abundance is usually measured as a good proxy of the metallicity in the interstellar medium of galaxies, since oxygen is the most abundant metal and the emission lines from the most important ionization stages of oxygen can be easily observed in optical. Reliable metallicity measurement of the ionized gas in galaxies requires the measurement of the electron temperature from the ratio of the auroral to the nebular emission lines, such as [OIII]$\lambda\lambda$5007,4959/[OIII]$\lambda$4363. However, it is difficult to detect the [OIII]$\lambda$4363 line, as it is intrinsically weak. This line is too weak to be observed in metal-rich environments (due to low electron temperature) or faint galaxies. When [OIII]$\lambda$4363 lines (or their analogs) are not detected,  metallicity-sensitive ratios of strong emission lines are widely used as metallicity indicators (strong-line methods), such as [NII]$\lambda$6584/H$\alpha$, ([OII]$\lambda\lambda$3726, 3729 + [OIII]$\lambda$$\lambda$4959,5007)/H$\beta$ (R23), [OIII]$\lambda$5007/[NII]$\lambda$6584, [SII]$\lambda\lambda$6716, 6731/H$\alpha$, [NII]$\lambda$6584/[SII]$\lambda\lambda$6716, 6731.   Strong-line methods are especially common in studies of high-redshift galaxies (e.g., Erb et al. 2006;  Mannucci et al.2010; Finkelstein et al. 2011; Belli et al. 2013; Henry et al. 2013; Kulas et al. 2013; Nakajima et al. 2013; Maier et al. 2014; Song et al. 2014; Steidel et al. 2014; Wuyts et al. 2014; Zahid et al. 2014; Sanders et al. 2015; Shapley et al. 2017). The strong line metallicity indicators have been typically calibrated in two ways: grids of photoionization models (McGaugh 1991; Zaritsky, Kennicutt $\&$ Huchra 1994; Kewley $\&$ Dopita 2002; Kobulnicky \& Kewley 2004; Tremonti et al. 2004; Dopita et al. 2013, 2016, etc); and samples of galaxies or HII regions for which the oxygen abundances have been well determined through the T$_e$ method (Pettini $\&$ Pagel 2004; Pilyugin $\&$ Thuan 2005; Yin et al. 2007; Pilyugin, Vılchez $\&$ Thuan 2010b; Pilyugin, Grebel $\&$ Mattsson 2012; Marino et al. 2013; Pilyugin $\&$ Grebel 2016; Curti et al. 2017, etc).

R23 is the most commonly used such strong line ratio, first proposed by Pagel et al. (1979). The R23 indicator could be used for both metal-poor galaxies (12+log(O/H) $<$ 8.5) and metal-rich galaxies (12+log(O/H) $\geq$ 8.5) (Pagel et al. 1979; Edmunds $\&$ Pagel 1984; Skillman et al. 1989; McGaugh 1991; Kobulnicky et al. 1999; Pilyugin 2000; Tremonti et al. 2004. etc). Recently, Maiolino et al. (2008) and Curti et al. (2017) provided R23 calibrations, based on a combination of both low-metallicity and high-metallicity nearby star-forming galaxies. However, the applicability of these calibrations to extreme emission-line galaxies, namely galaxies with unusually large equivalent widths of high-excitation emission lines, is unclear. The physical properties (e.g. sizes, stellar masses, metallicities, sSFR, dust, ionization conditions) within most nearby galaxies are significantly different from those within extreme emission-line galaxies (e.g. Kniazev et al. 2004; Cardamone et al. 2009; Atek et al. 2011; Izotov etal. 2011; van der Wel et al. 2011; Maseda et al. 2014; Amorin et al. 2014, 2015; Yang et al. 2016; Yang et al. 2017). In fact, the physical properties of extreme emission-line galaxies resemble those within Lyman-alpha emitting galaxies at high-redshift (e.g. Cowie et al. 2011; Finkelstein et al. 2011; Smit et al. 2014; Amorin et al. 2015; Yang et al. 2016; Yang et al. 2017; Stark et al. 2017). In particular, among the extreme emission-line galaxies, green pea galaxies are known as best nearby analogs of high-redshift Ly$\alpha$ emitting galaxies found so far (Henry et al. 2015; Yang et al. 2016; Yang et al. 2017). An R23 calibration derived from a systematic dataset of nearby extreme emission-line galaxies should potentially be appropriate for high-redshift Lyman-alpha emitting galaxies and other high-redshift extreme emission-line galaxies.

Green pea galaxies looked green and appeared to be unresolved round point sources in Sloan Digital Sky Survey (SDSS) gri composite color image (Cardamone et al. 2009). Cardamone et al.(2009) systematically selected 251 green peas from the SDSS Data Release 7 (DR7) by their photometric color criteria. Only 80 of these 251 are star-forming objects with high S/N SDSS spectra, and they are in the relatively narrow redshift range $0.14 < z < 0.36$. The key properties of these green peas are the compact sizes and large [OIII]$\lambda$5007 equivalent widths (300 - 2500\AA). In this paper, we select a considerably larger systematic dataset of $\sim$ 800 green pea galaxies from the spectroscopic database of SDSS Data Release 13 (Albareti et al. 2017). We derive a new empirical calibration of R23 for extreme emission-line galaxies using this systematic dataset of green pea galaxies. By combining R23 with [OIII]$\lambda$$\lambda$4959,5007/[OII]$\lambda$$\lambda$3726,3729 (hereafter ``[OIII]/[OII]"), our new calibration breaks the double-value degeneracy of R23 with metallicities in the regime of log [OIII]/[OII] $\geq$ 0.6. We also compare our calibration with previous calibrations. 

\section{Sample selection}
Our sample of green pea galaxies was selected from SDSS Data Release 13. The sample selection details and a full description of the sample are in Yang et al. in preparation. The sample selection steps are as follows.

1. The sample was pre-selected from ``galSpecLine" catalog by the MPA-JHU group (Brinchmann et al. 2004, Kauffmann et al. 2003, and Tremonti et al. 2004) in SDSS Data Release 8 and ``emissionLinesPort" catalog by Portsmouth Group (Thomas et al. 2013) in SDSS Data Release 12. Both catalogs contain emission line fluxes measurements for galaxy spectra. In each catalog, the criteria are:

\hangindent=0.76cm a) The spectroscopic classification of the object is ``Galaxy," and its subclass is consistent with a green pea galaxy--- that is, the subclass is ``starforming'' or ``starburst'', or ``NULL'', but not ``AGN''.

\hangindent=0.76cm b) The [OIII]$\lambda$5007 and H$\beta$ lines are well detected, with signal-to-noise ratio of the emission lines [OIII]$\lambda$5007 and H$\beta$ is greater than 5.

\hangindent=0.76cm c) The lines are strong: either the equivalent width of [OIII]$\lambda$5007 is EW([OIII]$\lambda$5007)$> 300$\AA, or the equivalent width of H$\beta$ is EW(H$\beta$)$> 100 $\AA. 

\hangindent=0.76cm d) The galaxy is spatially compact: petroR90$\textunderscore$r is smaller than 3.0$\arcsec$. petroR90\textunderscore r is the radius containing 90$\%$ of Petrosian flux in SDSS r band.

The union of the objects selected from both catalogs gives 1119 objects.

2. Note that ``galSpecLine" catalog is available for Data Release 8 galaxies and that ``emissionLinesPort" catalog reported an emission line measurement only when the amplitude-over-noise ratio is larger than two. We took the SDSS Data Release 13 pipeline results for the following selection and data analysis. We selected galaxies for which the fluxes of  [OII]$\lambda$3726, [OII]$\lambda$3729, H$\beta$, [OIII]$\lambda$5007,  H$\alpha$, and the corresponding flux uncertainties are all positive numbers. 69 objects that are classified as either AGNs or LINERs in the BPT diagram (Baldwin et al. 1981) by two classification lines proposed by Kewley et al. (2001) and Kauffmann et al. (2003) were excluded. 1004 objects were identified as star-forming galaxies. Note that the detection of [NII]$\lambda$6583 is not required in our sample selection. The objects with no detected [NII]$\lambda$6583 line are included in this work. Thus our sample is not biased toward high metallicity due to the [NII]$\lambda$6583 line.

3. Only the galaxies with signal-to-noise ratio of [OIII]$\lambda$4363 greater than 3 were selected. This criterion allows us to measure the metallicity with the T$_e$ method.     

After steps 1--3, we obtained a total of 835 galaxies, and these are our parent sample. The emission lines used in R23 measurements are all stronger than [OIII]$\lambda$4363. The [OII]$\lambda$3726 and [OII]$\lambda$3729 lines are typically the weakest of these for the present sample, but even they have a median S/N around 40, and always have S/N $> 4$ even in the cases of very high ionization. The size of our sample is ten times larger than that of the original spectroscopic sample of star forming green pea galaxies in Cardamone et al.(2009). Our sample covers the redshift range 0.011 $<$ z $<$ 0.411, as shown in Figure 1. We corrected the emission line fluxes for dust extinction using the Balmer decrement measurements. Assuming that the hydrogen lines emit from an optically thick HII region obeying Case B recombination, we took the intrinsic H$\alpha$/H$\beta$ ratio of 2.86. We adopted Calzetti et al. (2000) extinction curve. Therefore the nebular color excess is 
\begin{equation} 
E(B - V)_{gas} = \frac{\log_{10}[(f_{H\alpha}/f_{H\beta})/2.86]}{ 0.4\times[k(H\beta) - k(H\alpha)]} ~~,
\end{equation}
where k(H$\alpha$) = 3.33 and  k(H$\beta$)= 4.6. E(B - V)$_{gas}$ for our galaxies is small, typically lower than 0.4 mag, with the median E(B - V)$_{gas}$ of 0.11 mag.

\section{T$_e$-method determination of metallicity}

To derive the electron temperature and metallicity, we used the relations in Izotov et al. (2006) section 3.1. This follows the approach of most T$_e$-based metallicity studies. In this approach, a two-zone HII region model with two different electron temperatures is assumed. We used extinction-corrected line fluxes when measuring metallicities. We summarize the steps here but more details can be found in Izotov et al. (2006). We estimated the O$^{++}$ electron temperature T$_e$([OIII]) from the flux ratio [OIII]$\lambda\lambda$5007,4959/[OIII]$\lambda$4363 using Equations 1 and 2 of Izotov et al. (2006), then we estimated the O$^{+}$ electron temperature by 
\begin{equation}
t_2 = - 0.577 + t_3\times(2.065 - 0.498\times t_3), 
\end{equation}
where t$_2$ = 10$^{-4}$T$_e$([OII]), t$_3$= 10$^{-4}$T$_e$([OIII]). This relation was from photoionization models and was found in Izotov et al. (2006) to be consistent with observations. Note that our measurement of the oxygen abundance depends little on the relation between t$_2$ and t$_3$. If we change T$_e$([OII]) by +2000K or --1000K, the oxygen abundance 12 + log (O/H) differs only by $<$0.04 dex. Since the dominant ions of oxygen in HII regions are O$^{+}$ and O$^{++}$, the oxygen abundances are O/H $\approx$ (O$^+$ + O$^{++}$)/H$^+$. For the measurement of O$^{+}$ and O$^{++}$ abundances, we used [OII]$\lambda$$\lambda$3726,3729/H$\beta$ and ([OIII]$\lambda$4959+[OIII]$\lambda$5007)/H$\beta$. The equations are:
\begin{equation}
12 + \log \frac{O^{+}}{H^{+}} = \log \frac{[OII]\lambda3726 + [OII]\lambda3729}{H\beta} + 5.961 + \frac{1.676}{t_2} - 0.40\log t_2 - 0.034t_2 + \log(1 + 1.35\times10^{-4}n_et_3^{-0.5})
\end{equation} 
and 
\begin{equation} 
12 + \log \frac{O^{++}}{H^{+}} = \log \frac{[OIII]\lambda4959 + [OIII]\lambda5007}{H\beta} + 6.200 + \frac{1.251}{t_3} - 0.55\log t_3 - 0.014\, t_3
\end{equation}

We measured electron density from the flux ratio R =[SII]$\lambda$6716/[SII]$\lambda$6731 for the objects that have signal-to-noise ratio of [SII]$\lambda$6716 and [SII]$\lambda$6731 greater than 2 (779 objects). If 0.51 $\leq$ R $\leq$ 1.43 (607 objects), then R is sensitive to n$_e$, and n$_e$ was derived from the fitted function 
\begin{equation}
R(n_e) = a\frac{b+n_e}{c+n_e}
\end{equation}
between n$_e$ and R over a range of electron densities of 10 cm$^{-3}$ to 10$^4$ cm$^{-3}$, based on the temden package in IRAF, with a = 0.4441, b = 2514, and c = 779.3. If R $<$ 0.51 (only one object), we assumed an electron density of 10$^4$ cm$^{-3}$.  If R $>$ 1.43 (171 objects), we assumed an electron density of 10$^{0.5}$ cm$^{-3}$. For the other objects that do not have good S/N of either [SII]$\lambda$6716 or [SII]$\lambda$6731, we assumed an electron density of 100 cm$^{-3}$ (56 objects). We note that the assumption of n$_e$ = 10, 100, or 10$^3$ cm$^{-3}$ gives nearly same results of T$_e$([OIII]) and oxygen abundances. 

Monte Carlo simulations were applied to estimate the uncertainties of the T$_e$-based metallicity measurement. For each object, we generated 1000 realizations of the fluxes of four emission lines that are involved in the metallicity measurement, [OIII]$\lambda$4363, [OIII]$\lambda$5007, [OII]$\lambda$$\lambda$3726,3729, H$\beta$. For each emission line, the 1000 realizations followed the normal distribution with $\sigma$ equal to the 1$\sigma$ uncertainty associated with the flux of that line. Therefore, for each object, there is a distribution of 1000 metallicity measurements from the simulations. The measurement that corresponds to the maximum probability is taken to be the reported metallicity measurement value. And the surrounding 68.27$\%$ confidence interval is taken to be the 1$\sigma$ uncertainty of measurement. Figure 2 shows the distribution of the metallicity measurements for four objects in our parent sample as examples. For the whole parent sample, the uncertainties of the O$^{++}$ electron temperature T$_e$([OIII]) are typically 200 -- 400 K, and the uncertainties of the metallicity O/H are typically 0.02 -- 0.10 dex. 

In our parent sample, the typical O$^{++}$ electron temperature T$_e$([OIII]) is 10000 -- 18000 K, and the range of metallicities is 7.2 $<$ 12+log(O/H) $<$ 8.6. 15 galaxies with metallcities lower than 1/12 solar (12+log(O/H) $<$ 7.6) are found in our parent sample. The lowest two metallicities are 12+log(O/H) = 7.25, 7.26. Extremely metal-poor galaxies are particularly interesting, as they provide a unique opportunity to study physical processes in conditions that are characteristic of the early universe, such as star formation in low metallicity environments.

The distribution of our parent sample in the parameter space R23 vs log(O/H) is presented in Figure 3. The objects with 1$\sigma$ metallicity uncertainties higher than 0.15 dex, or with 1$\sigma$ R23 uncertainties higher than 0.02 dex, are shown with a reddish color, and their uncertainties are shown with error bars. These objects (5.5$\%$ of the parent sample) were excluded from our calibration of R23, leaving 789 objects with small uncertainties for that calibration.


\section{R23 calibration}

The R23 ratio depends on both the oxygen abundance and the physical conditions, as characterized, for example, by the hardness of the ionizing radiation or ionization parameter of HII regions. Adding [OIII]/[OII] as an additional parameter in the calibration of R23 indicator has been proposed (McGaugh 1991; Kobulnicky et al. 1999; Kewley \& Dopita 2002), since [OIII]/[OII] has a strong dependence on the ionization parameter, and the combination of [OIII]/[OII] with R23 can potentially separate the effects of ionization parameter and oxygen abundance. Similarly, Pilyugin (2000, 2001a,b) added p2 = log [OII]$\lambda$3726,3729/H$\beta$ - log R23 and p3 = log [OIII]$\lambda$$\lambda$4959,5007/H$\beta$ - log R23 in the calibration of R23 -- (O/H) relation, in order to separate the effects of ionization parameter.

We plot our sample in R23 vs log(O/H) parameter space again in Figure 4. We plot objects in the different ranges of log [OIII]/[OII] in different panels. As we can see, the separation of objects by [OIII]/[OII] largely decreases the scatter of objects. This is also seen in Figure 5, where the data points in the parameter space R23 vs 12+log(O/H) color-coded by [OIII]/[OII] are presented in a single panel.

In this work, we calibrated R23 with the parameter [OIII]/[OII]. When performing least squares fitting to the 789 objects, we applied the functional form

\begin{equation}
\log R23 = a + b\times x + c \times x^2 - d\times  (e + x) \times y,
\label{eq:r_of_z}
\end{equation}
where $x$ = 12+log(O/H) and $y$ = log [OIII]/[OII].
The functional form is new to this work. It is inspired by two functional forms in the literature. The first is the second-order polynomial function log $R23 = a + b\times x + c\times x^2$ with $x = 12+\log(O/H)$, which is used in R23 calibration studies such as Maiolino et al. (2008). The second is Equation 8 in Kobulnicky et al. (1999), which has the form $12+\log(O/H) = \alpha + \beta\times r + \gamma\times r^2 - y\times(\delta + \epsilon r + \zeta r^2)$  with r = log R23 and y = log [OIII]/[OII]. 

Since we do not know which data points are on the lower branch and which ones are on the upper branch, we fit for R23 as a function of metallicity and [OIII]/[OII] (i.e., R23 on the left side and metallicity and [OIII]/[OII] on the right side) instead of directly fitting for 12+log(O/H) as a function of R23 and [OIII]/[OII].   We begin with the ``traditional'' quadratic form, which we augment with a term $-y\times(de + dx)$ that incorporates $y=\log([OIII]/[OII])$ in a manner inspired by the approach of Kobulnicky (1999).

The coefficients of the best fit are 
\[
a = -24.135,  b = 6.1532,  c = -0.37866,  d = -0.147,  e = -7.071.
\]
If we apply S/N $>$ 5 in the [OIII]$\lambda$4363 line instead of S/N $>$ 3 when we selected the sample, the coefficients of the best fit would be a = -24.691, b = 6.3027, c = -0.38856, d = -0.146, e = -7.110. The R23 vs 12+log(O/H) distribution for the data points and these coefficients are similar no matter whether we apply S/N $>$ 5 in the [OIII]$\lambda$4363 line or S/N $>$ 3.
 
Our best fit is shown in Figure 4. According to the analytic expression of the best fit, when $\log([OIII]/[OII])$ changes, the relation between $\log R23$ and $12+\log(O/H)$ shifts. In Figure 4, the solid lines, from left to right and from top to bottom, show the curves of the best fit corresponding to log [OIII]/[OII] = 0.1, 0.35, 0.45, 0.55, 0.65, 0.75, 0.85, 0.95, 1.05, 1.15, 1.25, and 1.5, respectively. This calibration applies to the metallicity range of 7.2 $<$ 12+log(O/H) $<$ 8.6. For the four panels in the first row, the objects are in the turnover region of R23 diagnostics with some scatter. Therefore, the relation between R23 and log(O/H) derived in this work, could be used to estimate metallicities for objects with 0.0 $<$ log[OIII]/[OII] $<$ 0.6, but should be used with caution. For the second and third row, R23 follows an almost monotonic trend with metallicity and the objects show very small scatter. The calibration can safely be used to estimate metallicities for objects with log [OIII]/[OII] $\geq$ 0.6. For these objects, when solving metallicity, the lower branch solution should be taken. The curves of the best fit that correspond to different [OIII]/[OII] are also shown in a single panel in Figure 5.
 
 Inverting equation~\ref{eq:r_of_z} to solve for metallicity, we find the  solutions 

 \begin{equation}
12+\log(O/H) =   \left\{
    \begin{array}{ll}
      { \frac{(d\times y - b) - \sqrt{(b-d\times y)^2 - 4c\times(a- d\times e\times y - \log R23 ) }}{2c} } & 
      \hbox{for } y > 0.6 \hbox{ and }R23 \le R23_{max}(y)\\
      & \\
      { \frac{(d\times y - b) \pm \sqrt{(b-d\times y)^2 - 4c\times(a- d\times e\times y - \log R23 ) }}{2c} } & 
      \hbox{for } y \le 0.6 \hbox{ and }R23 \le R23_{max}(y) \\
      & \\
      { \frac{d\times y - b}{2c} } & 
      \hbox{for } R23 > R23_{max}(y)\\
    \end{array}
  \right.
  \label{eq:z_of_r}
\end{equation}

Here, again, $y\equiv \log([OIII]/[OII])$, and the coefficients a--e are given above.  When $\log([OIII]/[OII])> 0.6$, we find that the lower branch of the metallicity-R23 relation is suitable for all galaxies in our sample.   For smaller values of $\log([OIII]/[OII])$, our metallicity solution is double valued, and a supplemental branch indicator is needed.  Finally, observed values of $\log(R23) > \log\left(R23_{max}(y)\right) = a - d\times e\times y - (b - d\times y)^2/(4c)$ exceed the maximum R23 produced by our model, and are assigned the maximum metallicity value consistent with the observed value of $y$.  For our best fitting coefficients, the maximum R23 simplifies to 
$\log\left(R23_{max}(y)\right) = 0.862 + 0.155 y - 0.0143 y^2$. Equation~\ref{eq:z_of_r} can be readily used to infer metallicities for large samples of galaxies with [OII], [OIII], and H$\beta$ flux measurements.

In order to show the accuracy of our derived calibration for the objects with log [OIII]/[OII] $\geq$ 0.6 in our sample, in Figure 6, we plot $\Delta log(O/H)$. $\Delta log(O/H)$ = log(O/H) (R23) - log(O/H) (T$_e$), which is the difference between log(O/H) measured from T$_e$ and log(O/H) predicted by our empirical R23 calibration. $\Delta log(O/H)$ is presented with [OIII]/[OII], R23 and T$_e$-based metallicity, in different panels. For most objects, $\Delta log(O/H)$ is within $\sim$ 0.2 dex and the standard deviation of $\Delta log(O/H)$ is 0.14 dex. We also note that, in the second panel, for the objects with log [OIII]/[OII] $\geq$ 1.2, $\Delta log(O/H)$ is within $\sim$ 0.1 dex.  Additionally, $\Delta log(O/H)$ does not correlate with either [OIII]/[OII] or T$_e$-based metallicity, but it correlates with R23. 

We only selected the objects with detected [OIII]$\lambda$4363 lines (S/N $>$ 3) when performing the R23 calibration. We next wished to examine whether this selection biased our sample towards low-metallicity objects. There would be additional 169 objects in our sample, if we ignore the selection criterion on [OIII]$\lambda$4363 line but keep the other criteria unchanged. One object out the 169 objects has no detected continuum around wavelength 4363 $\AA$. For the other 168 objects, we estimated the 3$\sigma$ upper limit of [OIII]$\lambda$4363 emission line fluxes from SDSS spectra and then estimated the 3$\sigma$ lower limit of 12+log(O/H) with T$_e$ method. We have found that the objects with no detected [OIII]$\lambda$4363 lines are generally consistent with the same relation between R23 and log(O/H). 

From our own R23 calibration, we can estimate the metallicities for the 168 objects with no detected [OIII]$\lambda$4363. We took the lower branch solution for objects with log [OIII]/[OII] $>$ 0.6, as recommended in equation~\ref{eq:z_of_r}.  Where the solution is double valued at log [OIII]/[OII] $\leq$ 0.6, we note (from Figure 4) that a majority of galaxies lie on the lower branch solution for 0.5 $\leq$ log [OIII]/[OII] $\leq$ 0.6, and a majority on the upper branch solution for log [OIII]/[OII] $<$ 0.5. Therefore, we assigned the upper branch when log [OIII]/[OII] $<$ 0.5, and the lower branch otherwise.

Remember that we have 835 objects in the parent sample (see the text in section 2). The histogram of the metallicities for these 835 objects and the histogram for the 168 objects are shown in Figure 7. 

In Figure 8, we plot the contours of the calibration-derived metallicities in the R23 vs [OIII]/[OII] 2-dimensional parameter space for the regime of log [OIII]/[OII] $\geq$ 0.6. The solid lines are the contours of 12+log(O/H), from 7.3 to 8.3. The black dots are the 474 objects with log [OIII]/[OII] $\geq$ 0.6. Figure 8 provides a direct way to convert R23 and [OIII]/[OII] to metallicities.

\section{Discussion}
\subsection{comparison with calibrations in literature}
We compare our calibration with previous calibrations in this section. For empirical calibrations, we take Grasshorn Gebhardt et al. (2016) and Jones et al. (2015). For photoionization models, we take Kobulnicky $\&$ Kewley (2004). We also take semi-empirical calibrations in Maiolino et al. (2008). Note that Grasshorn Gebhardt et al. (2016), Jones et al. (2015), Maiolino et al. (2008) all used the approach of estimating direct metallicities in Izotov et al. (2006), which are directly comparable to our work.

We plot the R23 -- log(O/H) relations in Grasshorn Gebhardt et al. (2016) (blue dot-dashed line), Jones et al. (2015) (purple dashed line), Maiolino et al. (2008) (red dashed line) and this work (grey dashed lines) together with our sample (green dots) in Figure 9. As clearly seen, for our galaxies with 12+log(O/H) lower than $\sim$ 8.0, R23 changes more quickly as a function of log(O/H) than indicated by the relations in Grasshorn Gebhardt et al. (2016) and Maiolino et al. (2008). The maximum value of R23 indicated by the relation in Maiolino et al. (2008) is also low compared to our galaxies. When log R23 $<$ 0.95, the relation in Jones et al. (2015) underestimates the metallicities at a fixed R23 for our galaxies with 12+log(O/H) either lower than $\sim$ 8.0 or higher than $\sim$ 8.1. It would be more consistent with our galaxies if the whole relation is shifted towards the direction of higher metallicities.   

Grasshorn Gebhardt et al. (2016) derived R23 calibration based on 272 ``local counterparts'' with T$_e$-based metallicities of their emission-line star-forming galaxies at 1.9 $<$ z $<$ 2.35. The local counterparts are SDSS galaxies that have H$\beta$ luminosities greater than L(H$\beta$) $>$ 3$\times$10$^{40}$ ergs$^{-1}$ and are matched in both SFR and stellar mass to their 1.9 $<$ z $<$ 2.35 objects. The majority of their counterparts has metallicities 7.9 $<$ 12+(O/H) $<$ 8.5, with only $\sim$15 objects with 12+(O/H) $<$ 7.9 and only $\sim$ 4 objects with 12+(O/H) $<$ 7.8. Our sample includes more low-metallicity objects: 139 objects with 12+(O/H) $<$ 7.9 and 75 objects with 12+(O/H) $<$ 7.8. Their counterparts sample includes $\sim$90 objects with log[OIII]/[OII] $>$ 0.5 and $\sim$12 objects with log[OIII]/[OII] $>$ 0.8; while our sample includes more high-excitation objects: 598 objects with log[OIII]/[OII] $>$ 0.5 and 253 objects with log[OIII]/[OII] $>$ 0.8.

Jones et al. (2015) reported R23 calibration based on a local sample of 113 galaxies with H$\beta$ flux larger than 10$^{-14}$ ergs$^{-1}$cm$^{-2}$ and T$_e$-based metallicities from Izotov et al. (2006). They also reported 32 z $\sim$ 0.8 star-forming galaxies in the DEEP2 Survey that have a combined signal-to-noise of [OIII]$\lambda$$\lambda$4959, 5007 $>$ 80 and T$_e$-based metallicity measurement. They found that their R23 calibration is consistent with the z $\sim$ 0.8 galaxies. The majority of their local comparison sample has metallicities 7.9 $<$ 12+(O/H) $<$ 8.5, with only $\sim$8 objects with 12+(O/H) $<$ 7.9 and only 3 objects with 12+(O/H) $<$ 7.8. Their local sample includes $\sim$25 objects with log [OIII]/[OII] $>$ 0.5 and $\sim$10 objects with log [OIII]/[OII] $>$ 0.8. We plot their z $\sim$ 0.8 objects (purple squares) in Figure 9 as well. Although the R23 calibration from Jones et al. (2015) is not consistent with our sample, the z $\sim$ 0.8 objects do populate a similar region to our sample in the R23 vs 12+log(O/H) parameter space. One prominent difference between the z $\sim$ 0.8 objects and our sample is that all the z $\sim$ 0.8 objects have less extreme R23 values, with log R23 $<$ 1.0.

Maiolino et al. (2008) combined T$_e$-based metallicity for 259 low-metallicity (12+(O/H) $<$ 8.3) galaxies from the Nagao et al. (2006) with metallicity estimation for high-metallicity (12+(O/H) $>$ 8.4) SDSS DR4 star-forming galaxies derived from theoretical models by Kewley $\&$ Dopita (2002) to obtain a calibration in a wide metallicity range. The low-metallicity sample from the Nagao et al. (2006) consists of the star-forming galaxies with detected [OIII]$\lambda$4363 from SDSS DR3 (Izotov et al. 2006) and from the literature by 2006. Many galaxies in this low-metallicity sample are not extreme emission-line galaxies, with EW(H$\beta$) of at least $\sim$80 galaxies lower than 50 $\AA$ (see Figure 12 in Izotov et al. 2006).

To summarize, the discrepancy between the relations in Grasshorn Gebhardt et al. (2016), Jones et al. (2015), Maiolino et al. (2008) and our galaxies, seen in Figure 9, could be primarily due to the different sample selection approaches and the different sample size in the low metallicities regime.

To quantitatively compare the calibrations and our sample, in the left panels of Figure 10, we show the histograms of the differences between the T$_e$-based metallicities and the metallcities predicted by the different calibrations for the subset of 474 objects with log [OIII]/[OII] $\geq$ 0.6. From top to bottom, the calibrations are from this work, Grasshorn Gebhardt et al. (2016), Jones et al. (2015), Maiolino et al. (2008) and Kobulnicky $\&$ Kewley (2004), respectively. Kobulnicky $\&$ Kewley (2004) used the stellar population synthesis and photoionization models from Kewley $\&$ Dopita (2002). In their method, the gas metallicity and ionization parameter are determined simultaneously using the two line ratios of R23 and [OIII]/[OII] from an iterative approach. We took the lower branch solutions in Kobulnicky $\&$ Kewley (2004). The black dashed lines are the reference line where $\Delta log(O/H)$ = 0.0. In each panel, the median $\Delta log(O/H)$ ($\Delta$) and the standard deviation ($\sigma$) is written in the upper left region. For this work, $\Delta$ is very close to zero, which indicates there is no systematic offset between the T$_e$-based metallicities and the metallicities predicted by our calibration. The $\sigma$ of $\Delta log(O/H)$ estimated from our calibration is as small as 0.14 dex. Among the calibrations in the other 4 panels, Maiolino et al. (2008) systematically underestimate the meatallicities by 0.02 dex, with the $\sigma$ of $\Delta log(O/H)$ of 0.14 dex. Grasshorn Gebhardt et al. (2016) and Jones et al. (2015) underestimate the meatallicities by 0.13 dex and 0.10 dex. Kobulnicky $\&$ Kewley (2004) overestimate the meatallicities by 0.32 dex. In addition, in the right panels of Figure 10, we present $\Delta log(O/H)$ from the different calibrations as a function of the T$_e$-based metallicities. In the low-metallicity regime (12+log(O/H) $<$ 7.9), the calibration in this work (in the top right panel of Figure 10) predicts metallicities much better (with the standard deviation of $\sigma$ = 0.13 dex) than the other calibrations. In the low-metallicity regime (12+log(O/H) $<$ 7.9), Grasshorn Gebhardt et al. (2016) and Jones et al. (2015) systematically underestimate the metallicities (by 0.16 dex and 0.07 dex); Kobulnicky $\&$ Kewley (2004) systematically overestimate the metallicities by 0.39 dex; Maiolino et al. (2008) give large scatter with the standard deviation of $\sigma$ = 0.18 dex.

It should be kept in mind that, the $\Delta log(O/H)$ for Maiolino et al. (2008), Grasshorn Gebhardt et al. (2016) and Jones et al. (2015) shown here are on the ideal premise that we know exactly whether each object is on the upper or lower branch of R23-log(O/H). The real accuracies of Maiolino et al. (2008), Grasshorn Gebhardt et al. (2016), Jones et al. (2015) may be not as good as the median and standard deviation values reported here.

\subsection{The applicability of R23 indicator at high redshift}
We highlight that green peas are best nearby analogs of high-redshift Ly$\alpha$ emitting galaxies. This suggests that our empirical calibration of R23 can be applied to high-redshift Ly$\alpha$ emitting galaxies. However, how about the applicability of our calibration to other star-forming galaxies (e.g. [OIII] emitters, H$\alpha$ emitters) at high redshift? 

In the [OIII]$\lambda$5007/H$\beta$ vs [NII]$\lambda$6584/H$\alpha$ BPT diagram, high-redshift galaxies have been found to be offset from the local SDSS galaxies (e.g. Steidel et al. 2014; Shapley et al. 2015). This raises concerns about estimating metallicities at high redshift from metallicity indicators based on nitrogen emission lines (e.g., the [NII]/H$\alpha$ and [OIII]/H$\beta$/([NII]/H$\alpha$) indicators).   
Among the common strong-line indicators, R23 and [OIII]/[OII] are only based on oxygen and hydrogen emission lines, which are more direct probes of the oxygen abundance compared with strong-line indicators that involve nitrogen or sulfur lines. Moreover, Nakajima et al.(2013) (see their Figure 7), Shapley et al. (2015) (see their Figure 4) and Strom et al. 2017 (see their Figure 8) point out that high-redshift star-forming galaxies occupy the same region of R23 vs [OIII]/[OII] parameter space as low-metallicity, low-mass SDSS star-forming galaxies, with no evidence for a systematic offset. Also remember that z $\sim$ 0.8 galaxies in Jones et al. (2015) follow consistent R23 -- log(O/H) parameter space as our galaxies (see text in section 5.1). Therefore, the empirical calibration of R23 abundance indicator based on our z $\sim$ 0.3 low-metallicity star-forming galaxy sample, could potentially be a good way to measure the metallicity for high-redshift star-forming galaxies that have similar R23, [OIII]/[OII], and EW([OIII]) to our galaxies. This has yet to be confirmed with direct T$_e$-based measurements of more high-redshift galaxies, though. We also emphasize that our calibration is only valid for the range of metallicities ($7.2 < 12 + \log(O/H) < 8.6$) and line ratios studied by this work.  Note also that [OIII]/[OII] is affected by dust extinction, and the use of this R23 indicator requires dust extinction correction.  The dust correction can be obtained from either Balmer decrement from H$\alpha$ and H$\beta$, or be estimated from SED fitting to broadband photometry or spectroscopy.  Empirically, the dust extinction is modest in our sample, and is likely to be similarly modest in other physically similar galaxy samples.

\section{Summary}

In this paper, we have assembled a large dataset of 835 star-forming green pea galaxies that spans a wide redshift range 0.011 $<$ z $<$ 0.411 from SDSS DR13. The main selection criteria are EW([OIII]$\lambda$5007) $>$ 300\AA\ or EW(H$\beta$) $>$ 100\AA\ and the S/N ratio of [OIII]$\lambda$4363 emission line higher than 3. We have measured electron temperature and T$_e$-based metallicities for these galaxies. The typical range of electron temperature is 10000 K - 18000 K. The metallicities vary from 7.2 to 8.6, with metallicities of 15 galaxies lower than 1/12 solar and the lowest metallicities being 12+log(O/H) = 7.25 and 7.26. 

We have derived new empirical calibration of the metallicities indicator R23 in strong line emitters based on 789 star-forming pea galaxies with a totally new functional form. Our calibration takes the analytic expression 
\[
\log R23 = a + b\times (12+\log(O/H)) + c \times (12+\log(O/H))^2 - d\times  (e + 12+\log(O/H)) \times \log [OIII]/[OII] 
\]
with coefficients 
\[
a = -24.135, b = 6.1532, c = -0.37866, d = -0.147, e = -7.071. 
\]
We have found that for objects with log [OIII]/[OII] $\geq$ 0.6, when separated by [OIII]/[OII], R23 shows an almost monotonic relation with 12+log(O/H) and there is no need to worry about the double-valued character of R23. Our calibration gives metallicity estimates that are accurate to within $\sim$ 0.14 dex in the regime of log [OIII]/[OII] $\geq$ 0.6. We also provide convenient equations (eq.~\ref{eq:z_of_r}) and plots (fig.~\ref{fig7}) to directly convert R23 and [OIII]/[OII] to metallicities. Our relations improve on prior work by reducing either bias or scatter for these extreme emission-line emitters.

Our sample galaxies are the best nearby analogs of high-redshift Lyman-alpha emitting galaxies, thus the calibration in this work could be very good for estimating the metallicities for high-redshift Ly$\alpha$ emitters from R23 and [OIII]/[OII]. Considering that R23 and [OIII]/[OII] only involve oxygen and hydrogen lines, and there is no evidence for a systematic offset between many high-redshift star-forming galaxies and the low-metallicity, low-mass SDSS star-forming galaxies in the R23 vs [OIII]/[OII] parameter space, this calibration could also be potentially applied to many other high-redshift star-forming galaxies.

\acknowledgments 
We thank the US National Science Foundation for its financial support through grant NSF AST-1518057, and NASA for its financial support through the WFIRST Preparatory Science program, grant number NNX15AJ79G. This work is based in part on observations made with the NASA/ESA Hubble Space Telescope, obtained [from the Data Archive] at the Space Telescope Science Institute, which is operated by the Association of Universities for Research in Astronomy, Inc., under NASA contract NAS 5-26555. These observations are associated with program \#14201. Support for program \#14201 was provided by NASA through a grant from the Space Telescope Science Institute, which is operated by the Association of Universities for Research in Astronomy, Inc., under NASA contract NAS 5-26555. Tianxing Jiang thanks the LSSTC Data Science Fellowship Program, her time as a Fellow has benefited this work. This work has made use of data from the Sloan Digital Sky Survey (SDSS). Funding for the Sloan Digital Sky Survey IV has been provided by the Alfred P. Sloan Foundation, the U.S. Department of Energy Office of Science, and the Participating Institutions. SDSS-IV acknowledges
support and resources from the Center for High-Performance Computing at
the University of Utah. The SDSS web site is www.sdss.org.SDSS-IV is managed by the Astrophysical Research Consortium for the 
Participating Institutions of the SDSS Collaboration including the 
Brazilian Participation Group, the Carnegie Institution for Science, 
Carnegie Mellon University, the Chilean Participation Group, the French Participation Group, Harvard-Smithsonian Center for Astrophysics, 
Instituto de Astrof\'isica de Canarias, The Johns Hopkins University, 
Kavli Institute for the Physics and Mathematics of the Universe (IPMU) / 
University of Tokyo, Lawrence Berkeley National Laboratory, 
Leibniz Institut f\"ur Astrophysik Potsdam (AIP),  
Max-Planck-Institut f\"ur Astronomie (MPIA Heidelberg), 
Max-Planck-Institut f\"ur Astrophysik (MPA Garching), 
Max-Planck-Institut f\"ur Extraterrestrische Physik (MPE), 
National Astronomical Observatories of China, New Mexico State University, 
New York University, University of Notre Dame, 
Observat\'ario Nacional / MCTI, The Ohio State University, 
Pennsylvania State University, Shanghai Astronomical Observatory, 
United Kingdom Participation Group,
Universidad Nacional Aut\'onoma de M\'exico, University of Arizona, 
University of Colorado Boulder, University of Oxford, University of Portsmouth, 
University of Utah, University of Virginia, University of Washington, University of Wisconsin, 
Vanderbilt University, and Yale University.

\clearpage
\begin{figure}
\includegraphics[scale=.70]{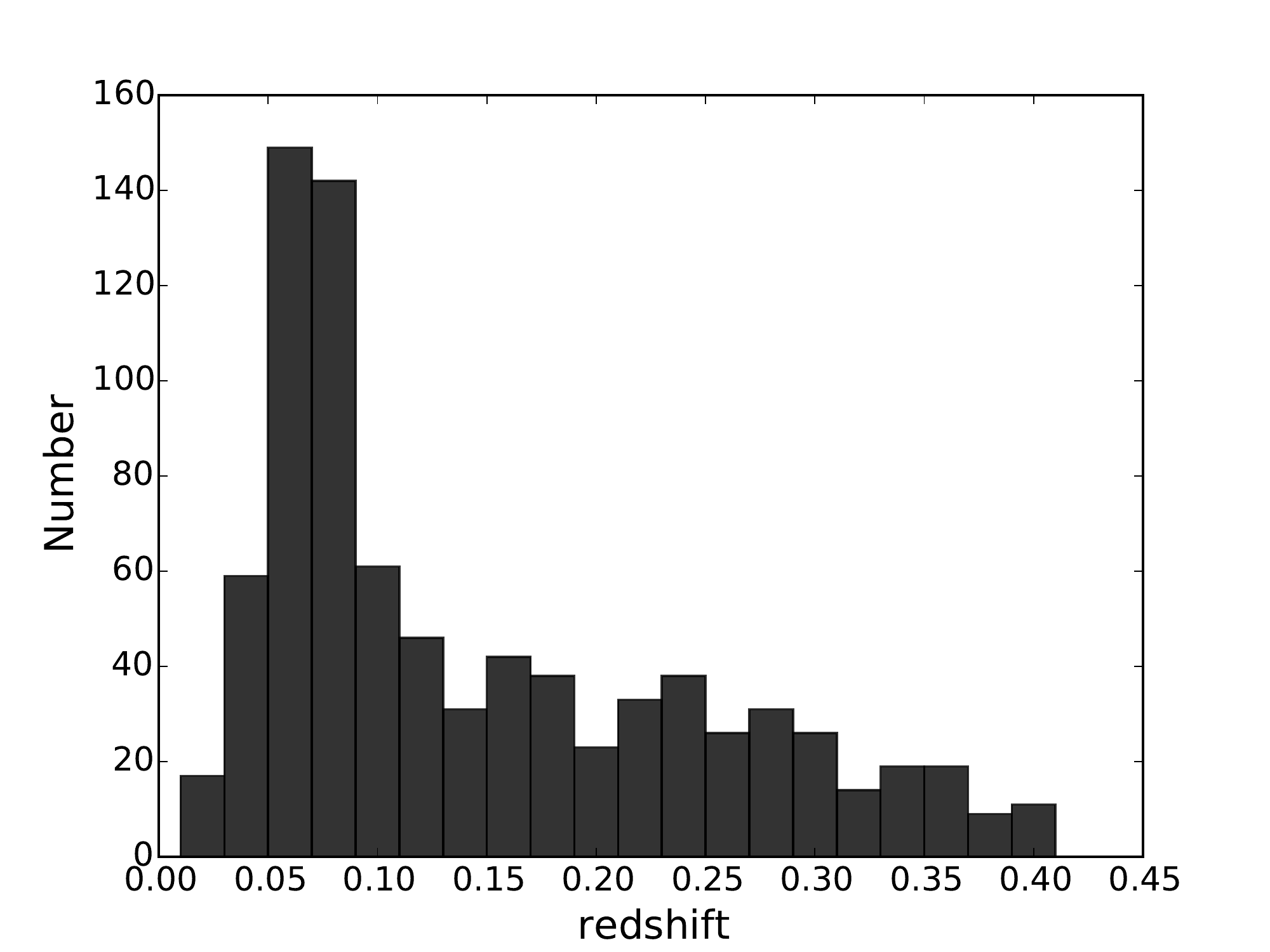}
\caption{The distribution of redshift for our parent sample of 835 galaxies.
}\label{fig1}
\end{figure}

\begin{figure}
\includegraphics[scale=.53]{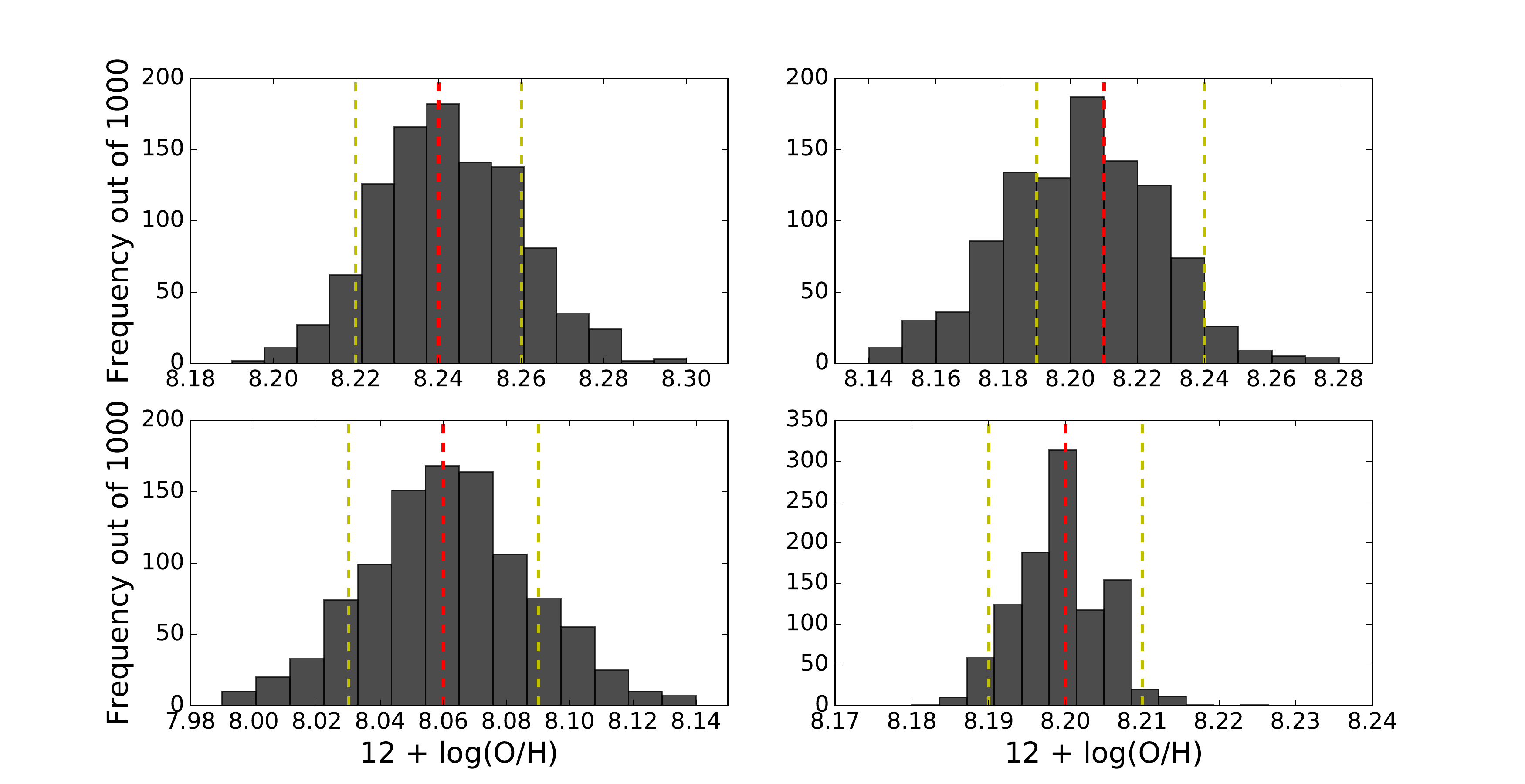}
\caption{The distribution of metallicity measurements from Monte Carlo simulations of line flux uncertainties for four objects in our parent sample, as examples. The red line shows the reported measurement value of the metallicity for this object. The yellow lines show the 68.27$\%$ confidence interval, which we use to derive the reported metallicity uncertainty. These four objects are randomly chosen from our sample.
}\label{fig2}
\end{figure}

\begin{figure}
\includegraphics[scale=.68]{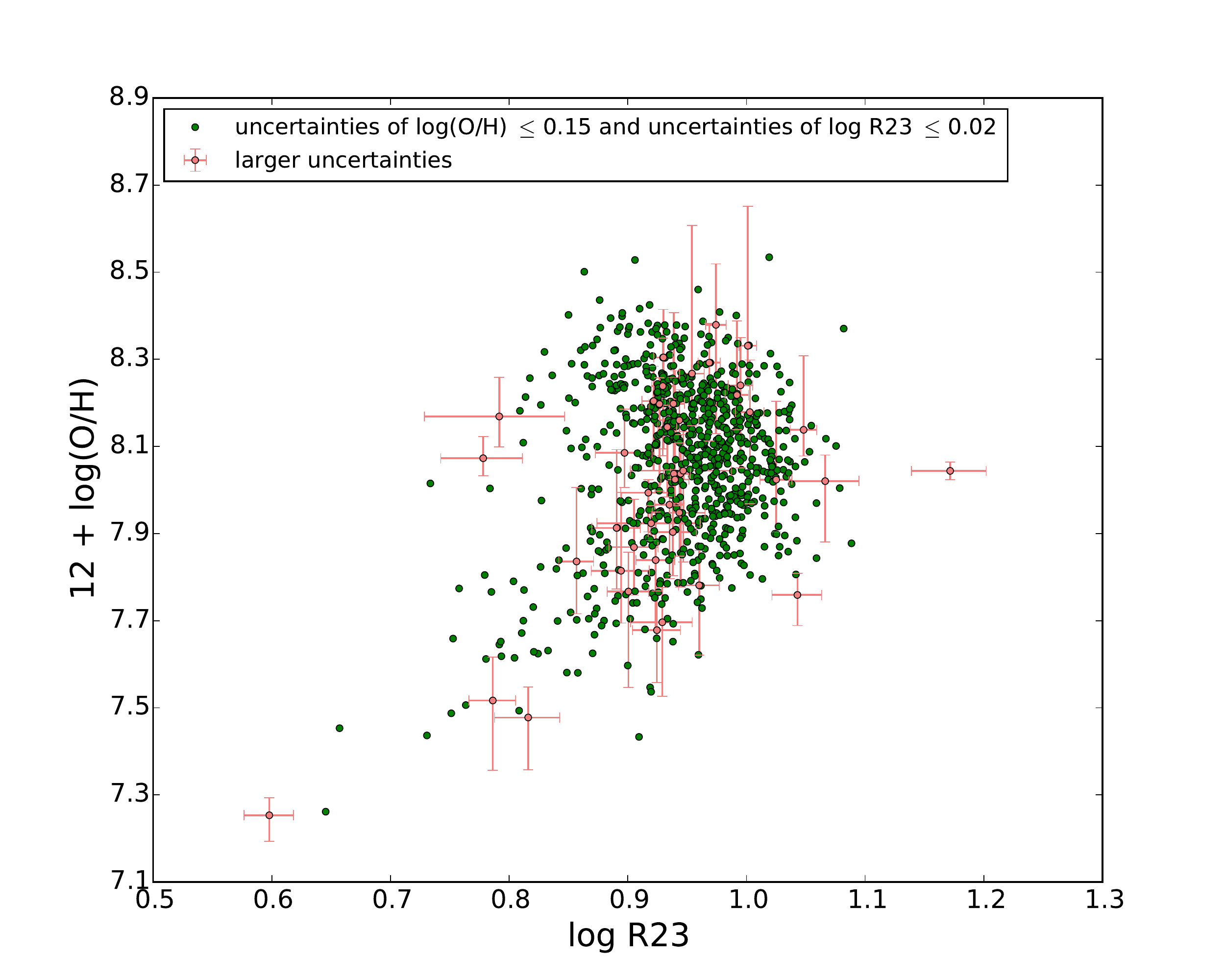}
\caption{log R23 vs 12 + log(O/H) for our parent sample. The green dots  (789 objects) are the objects with uncertainties no greater than 0.15 dex on O/H (as derived from the T$_e$ method), and uncertainties no greater than 0.02 dex on R23. The reddish dots with error bars are the objects that do not satisfy these uncertainty criteria.  These objects were excluded in the R23 calibration work. We have found two objects with 12+log(O/H) $<$ 7.3 in our parent sample (the two objects in the bottom left corner).
}\label{fig3}
\end{figure}

\begin{figure}
\includegraphics[scale=.557]{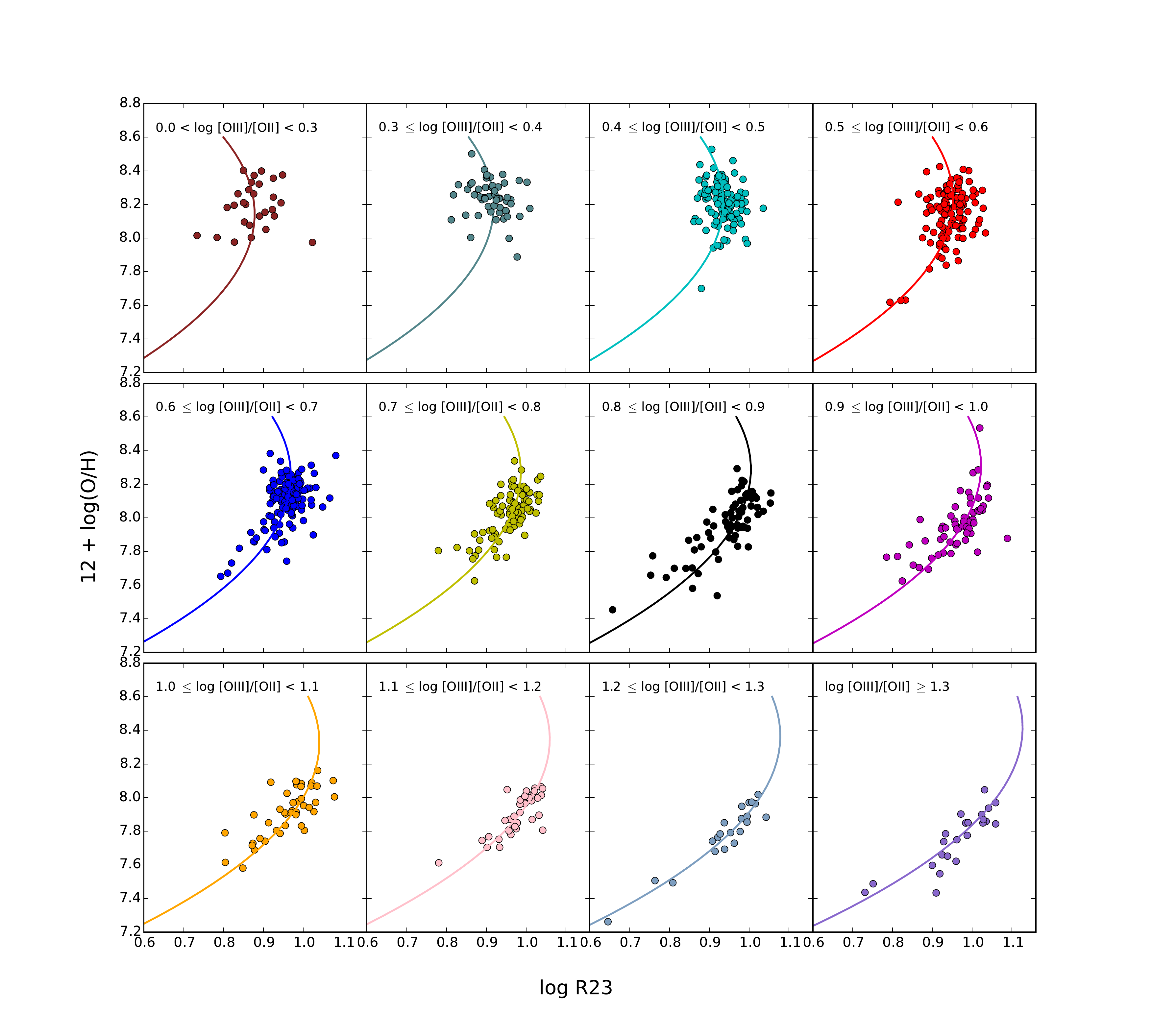}
\caption{The filled circles show our sample (789 objects) in the parameter space log R23 vs 12 + log(O/H). The objects in different ranges of log [OIII]/[OII] are separated into different panels. We did least squares fitting to the 789 objects by applying the functional form log R23 = a + b$\times$ (12+log(O/H)) + c $\times$ (12+log(O/H))$^2$ - d$\times$  (e + 12+log(O/H)) $\times$ log [OIII]/[OII]. The solid lines, from left to right and from top to bottom, show the curves of the best fit when log [OIII]/[OII] = 0.1, 0.35, 0.45, 0.55, 0.65, 0.75, 0.85, 0.95, 1.05, 1.15, 1.25, 1.5, respectively. The solid lines are consistent with the data points in each panel, demonstrating the reliability of the fit between R23, [OIII]/[OII], and 12+log(O/H). Please refer to Section 4 for details.}\label{fig4}
\end{figure}

\begin{figure}
\includegraphics[scale=.81]{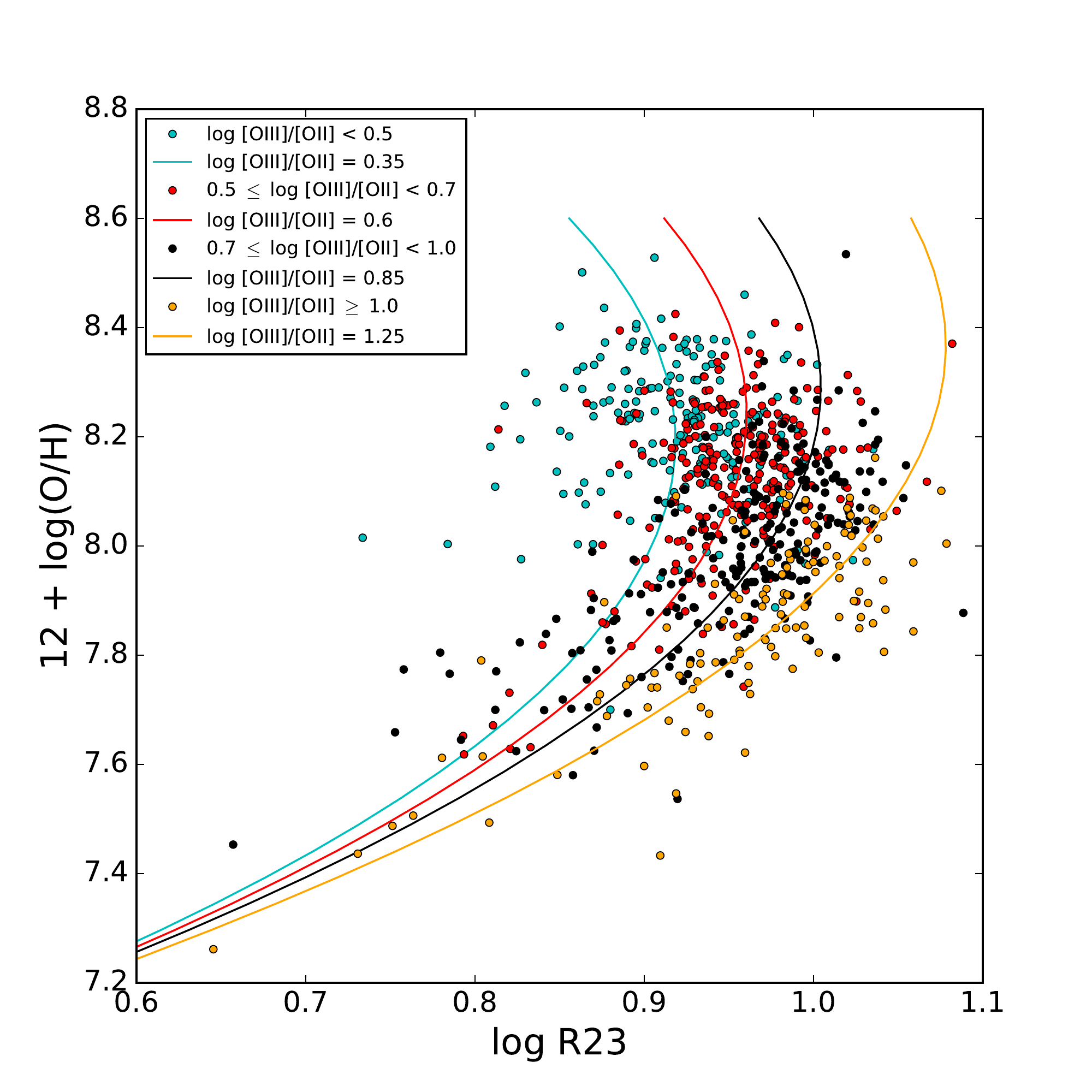}
\caption{Our sample (789 objects) in the parameter space log R23 vs 12 + log(O/H) color-coded by [OIII]/[OII] in a single panel. The solid lines, from left to right, show the curves of the best fit when log [OIII]/[OII] = 0.35, 0.6, 0.85, 1.25, respectively. This plot is to show that the data with different [OIII]/[OII] occupy different regions of the parameter space log R23 vs 12 + log(O/H) and to directly show the relative locations of the curves of the best fit corresponding to different [OIII]/[OII].}\label{fig5}
\end{figure}

\begin{figure}
\includegraphics[scale=0.408]{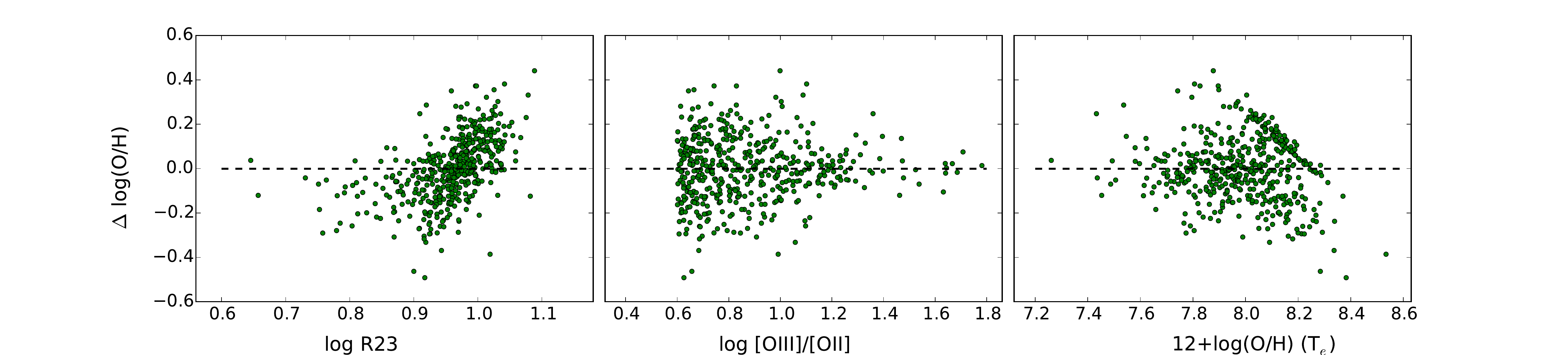}
\caption{Y axis: $\Delta log(O/H)$ = log(O/H) (R23) - log(O/H) (T$_e$), which is the difference between log(O/H) measured from T$_e$ and log(O/H) predicted by our empirical R23 calibration. X axis: log R23, log [OIII]/[OII], and metallicities measured from T$_e$. This is for the subset of the sample with log [OIII]/[OII] $\geq$ 0.6 (474 objects). There is a correlation between log R23 and $\Delta log(O/H)$. For most objects, the difference on the y axis is within $\sim$ 0.2 dex.}\label{fig6}
\end{figure}

\begin{figure}
\includegraphics[scale=.56]{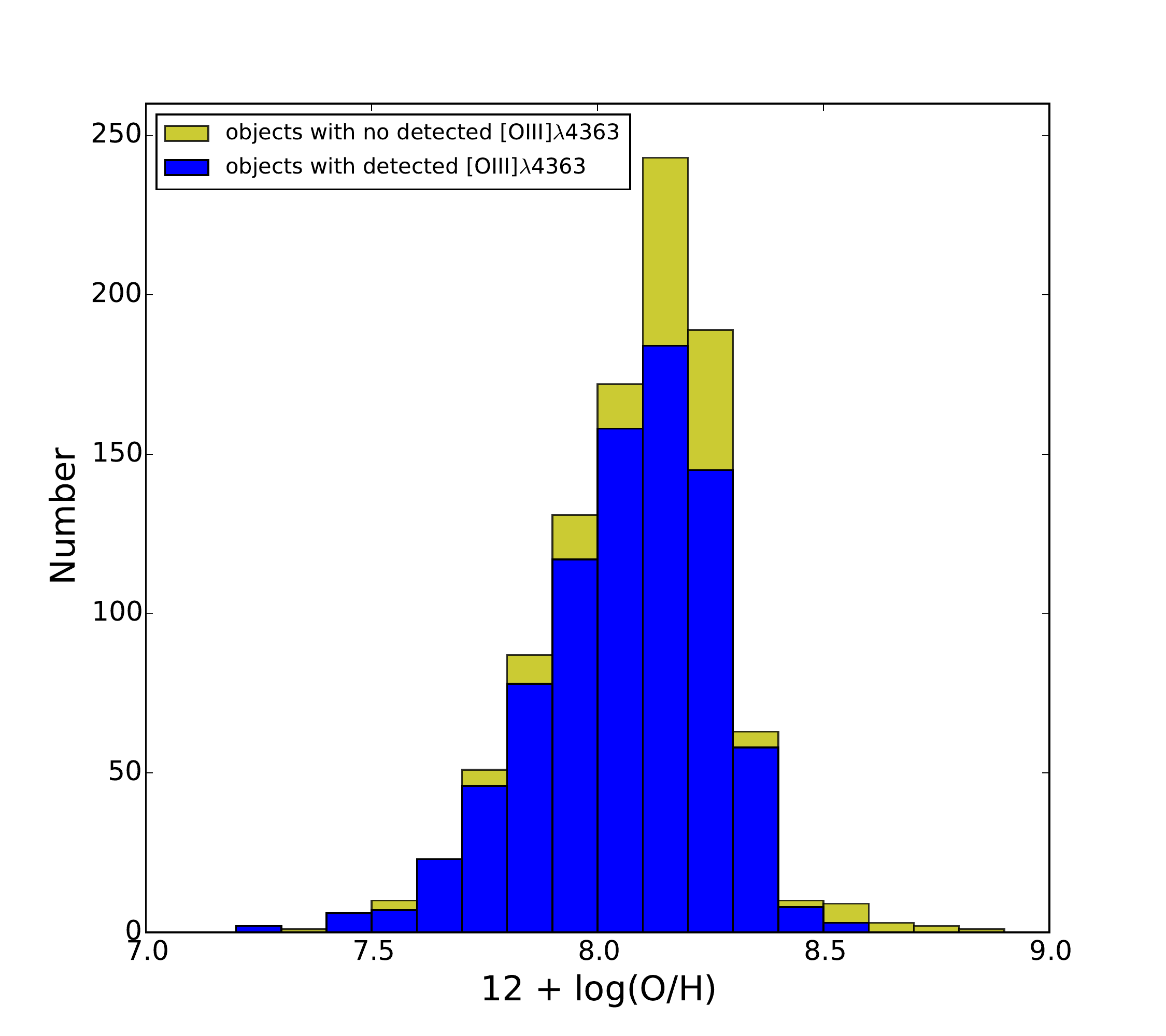}
\caption{Histogram of metallicities. The blue color shows $T_e$ based metallicities for our parent sample (refer to figure 3 for ``parent sample"). The yellow color represents the 168 objects with S/N of [OIII]$\lambda$4363 no greater than 3. The metallicities of these 168 objects are estimated from our own R23 calibration, using the lower branch for ratios log [OIII]/[OII]$>$0.5 and the upper branch for log [OIII]/[OII]$<$0.5 .}\label{fig7}
\end{figure}

\begin{figure}
\includegraphics[scale=.70]{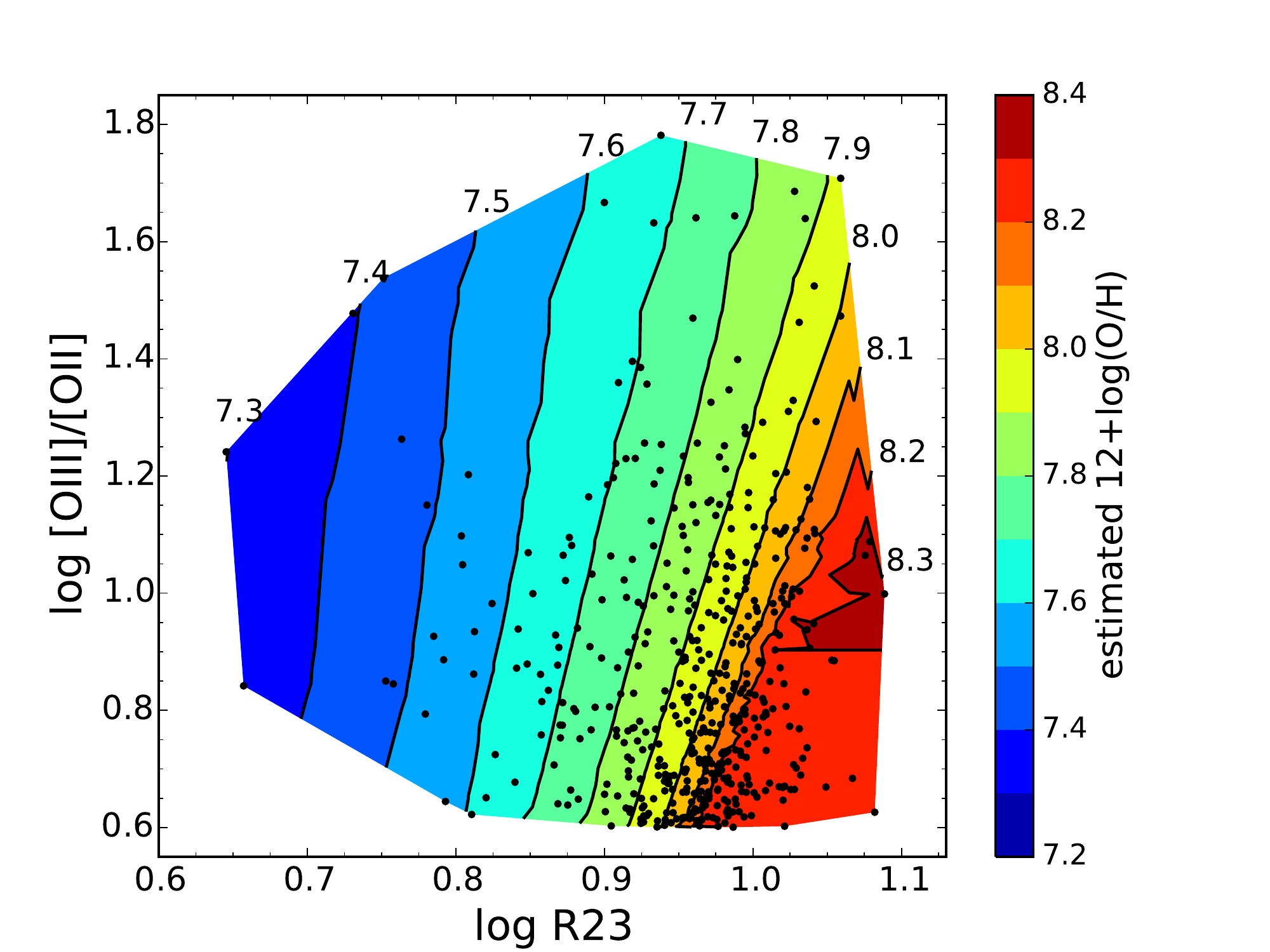}
\caption{Metallicity as a function of R23 and [OIII]/[OII] based on our R23 calibration in the regime of log [OIII]/[OII] $\geq$ 0.6. The black dots are a subset of the sample with log [OIII]/[OII] $\geq$ 0.6 (474 objects). The contours are drawn based on the metallicities of these dots that are estimated from our R23 calibration. This figure provides a direct way to estimate metallicities from R23 and [OIII]/[OII].}\label{fig8}
\end{figure}

\begin{figure}
\includegraphics[scale=.66]{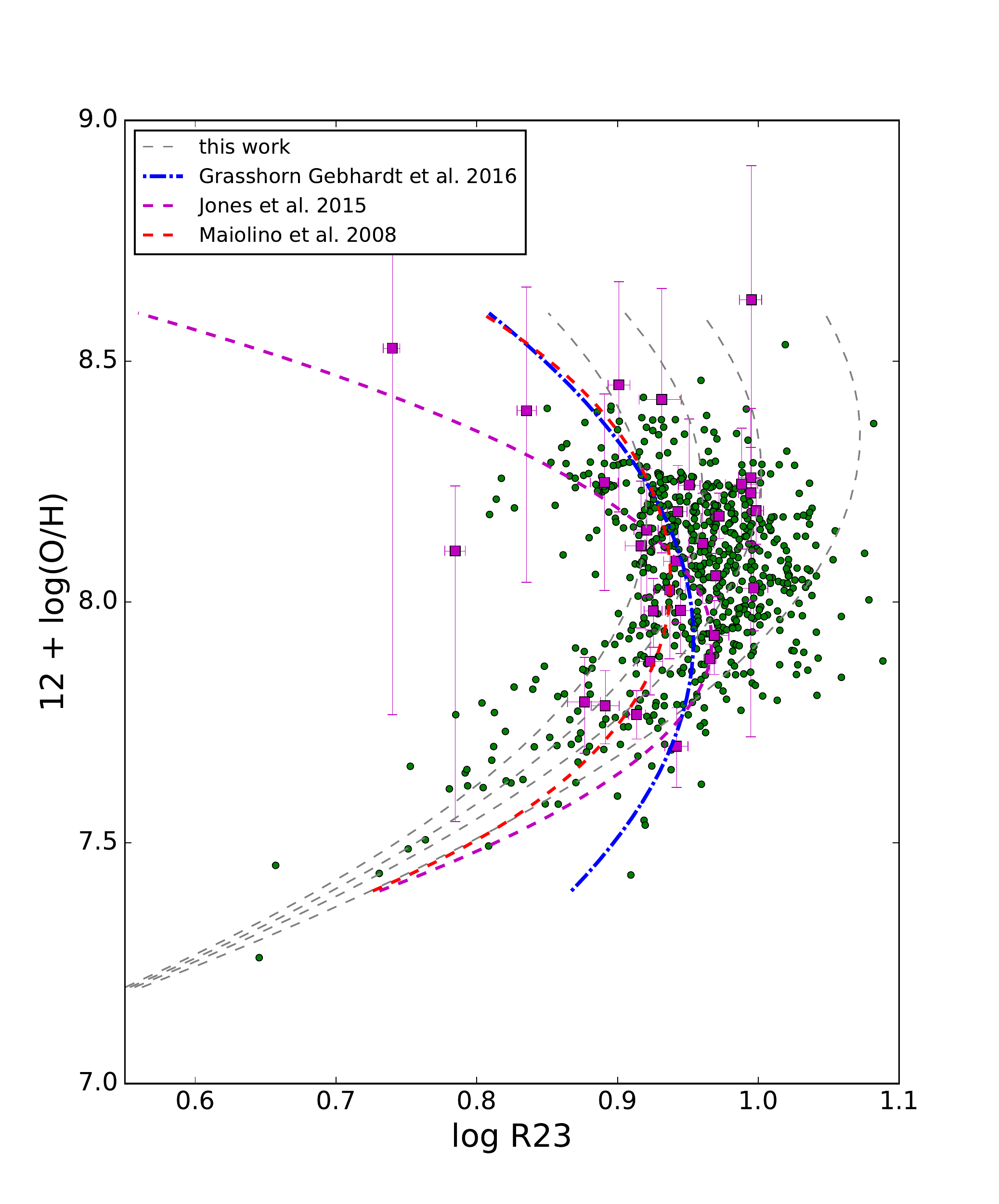}
\caption{The comparison between our sample (green dots), the calibration in this work (grey dashed lines), in Grasshorn Gebhardt et al. (2016) (blue dot-dashed line), in Jones et al. (2015) (purple dashed line), and in Maiolino et al. (2008) (red dashed line). The grey dashed lines, from left to right, show the curves of our calibration when log [OIII]/[OII] = 0.35, 0.6, 0.85, 1.25, respectively (same as the lines in Figure 5). The purple squares are the star-forming galaxies at z$\sim$0.8 in Jones et al. (2015). These galaxies lie in a similar region of parameter space as our sample. The calibration in Grasshorn Gebhardt et al. (2016) was based on the ``local counterparts'' of their 256 emission-line star-forming galaxies at z $\sim$ 2. The R23 calibration in Jones et al. (2015) was directly derived from their local comparison sample of 113 galaxies. The calibration in Maiolino et al. (2008) is derived from the combination of low-metallicity sample from Nagao et al. 2006 and high-metallicity star forming galaxies in SDSS DR4. All three calibrations from the literature show noticeable differences from our sample.}\label{fig9}
\end{figure}

\begin{figure}[ht] 
  \begin{subfigure}[b]{0.44\linewidth}
    \centering
    \includegraphics[scale = 0.56]{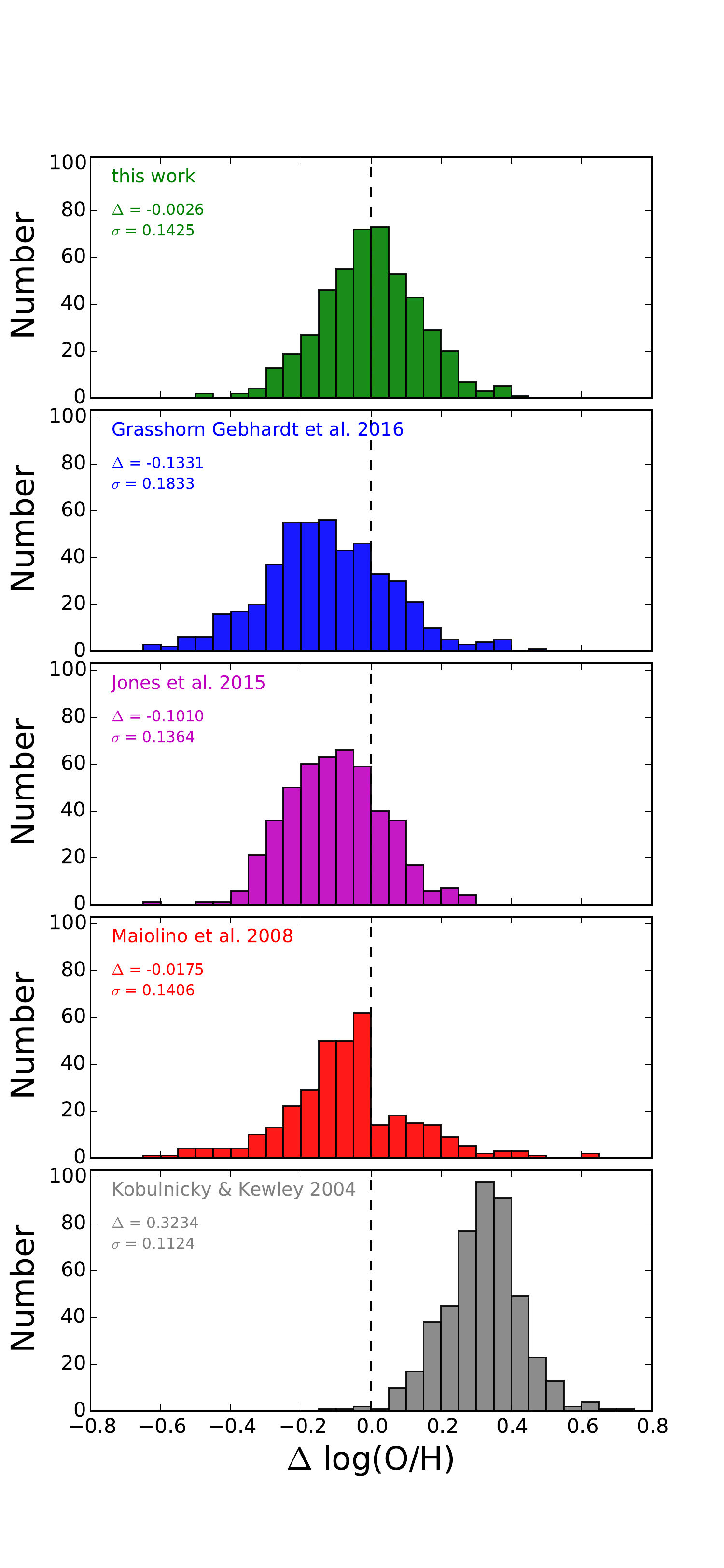} 
  
    \label{fig7:a} 

   \vspace{0.01cm}
  \end{subfigure} 
  \begin{subfigure}[b]{0.5\linewidth}
    \centering
    \includegraphics[scale = 0.56]{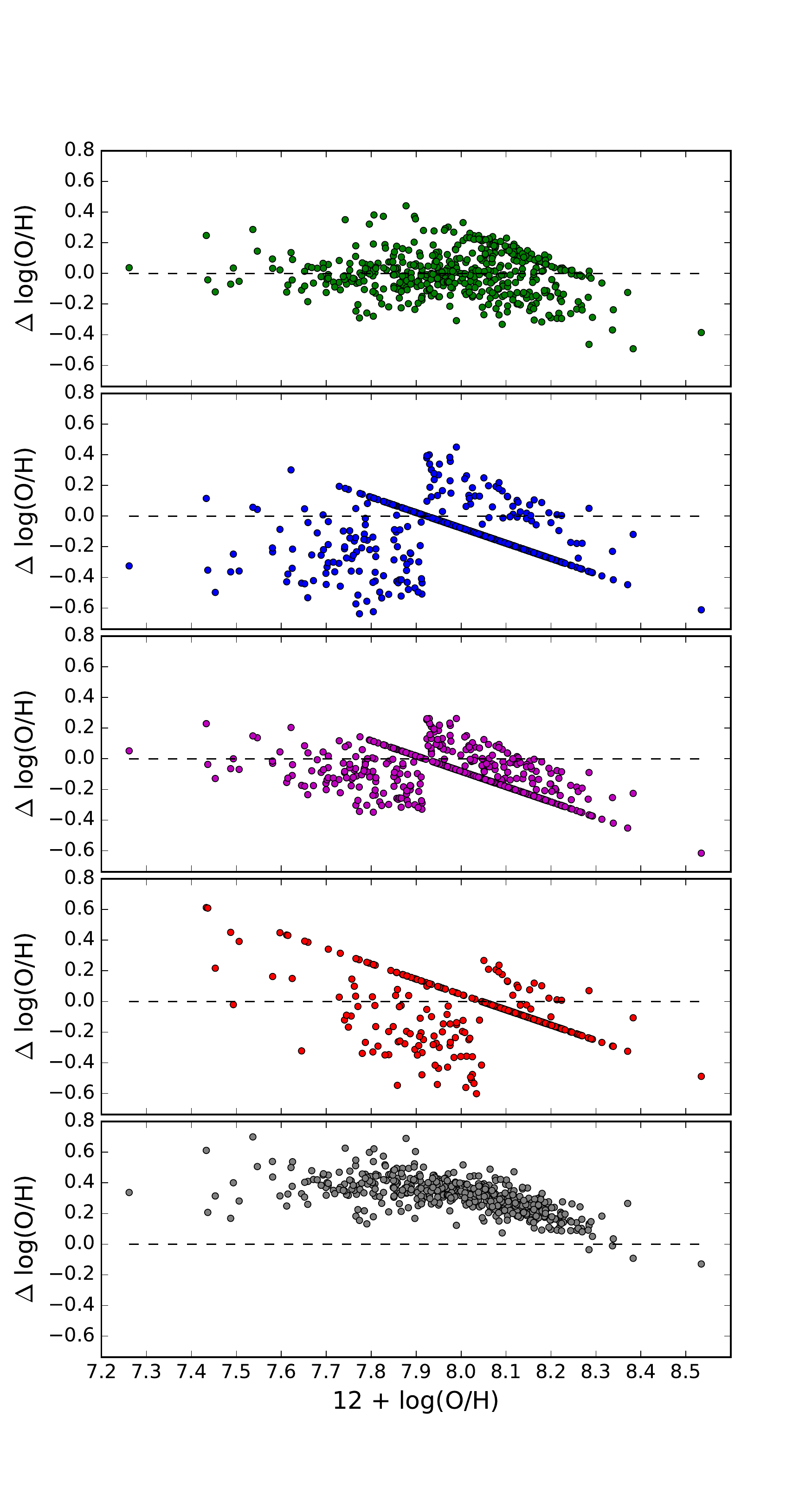} 
  
    \label{fig7:b} 
  \end{subfigure} 
  \caption{Left panels: Histograms of $\Delta$ log(O/H) for the subset of our sample with log [OIII]/[OII] $\geq$ 0.6 (474 objects). $\Delta$ log(O/H) = log(O/H) (R23) - log(O/H) (T$_e$), which is the difference between log(O/H) measured from T$_e$ and log(O/H) predicted by R23 calibrations. In different panels, the R23 calibrations are from this work, Grasshorn Gebhardt et al. (2016), Jones et al. (2015), Maiolino et al. (2008), and Kobulnicky $\&$ Kewley (2004), respectively. In each panel, the median $\Delta log(O/H)$ ($\Delta$) and the standard deviation ($\sigma$) is written in the upper left region. Grasshorn Gebhardt et al. (2016) and Jones et al. (2015) systematically underestimate the metallicities and Kobulnicky $\&$ Kewley (2004) systematically overestimate the metallicities. Right panels: $\Delta$ log(O/H) for the subset of our sample with log [OIII]/[OII] $\geq$ 0.6 (474 objects) vs 12+log(O/H) derived from T$_e$ method. In the low-metallicity regime (12+log(O/H) $<$ 7.9), the calibration in this work (in the top right panel) predicts metallicities much better than the other calibrations shown in the other 4 panels.  The diagonal feature visible in most panels corresponds to objects whose observed R23 value exceeds the maximum permitted by the model considered in that panel.  Such galaxies are all assigned the metallicity corresponding to the maximum allowed R23, and their residuals therefore fall on a line with $\Delta \log(O/H) = \log(O/H) (R23_{max}) - \log(O/H)(T_e)$.
}\label{fig10}
\end{figure}

\clearpage

\end{document}